\documentclass[letter]{aa}

\usepackage{graphicx}
\usepackage{txfonts}
\usepackage{natbib}
\usepackage{bbm}
\usepackage[breaklinks=true]{hyperref}
\hypersetup{colorlinks=true,urlcolor=blue,citecolor=blue,pdfborder= 0 0 0}
\bibpunct{(}{)}{;}{a}{}{,}             %% natbib format for A&A and ApJ
\begin{document}

   \title{Galaxy pairs align with galactic filaments}

   %\subtitle{None}

   \author{E.~Tempel\inst{1}
          \and A.~Tamm\inst{1}
          }

   \institute{Tartu Observatory, Observatooriumi~1, 61602 T\~oravere, Estonia\\
              \email{elmo.tempel@to.ee} }

   %\date{Received September 15, 1996; accepted March 16, 1997}
   \date{}

% \abstract{}{}{}{}{} 
% 5 {} token are mandatory
 
  \abstract
   {Gravitational collapse theory and numerical simulations suggest that the velocity field within large-scale galaxy filaments is dominated by motions along the filaments.}
   {Our aim is to check whether observational data reveal any preferred orientation of galaxy pairs with respect to the underlying filaments as a result of the expectedly anisotropic velocity field.}
   {We use galaxy pairs and galaxy filaments identified from the Sloan Digital Sky Survey data. For filament extraction, we use the Bisous model that is based the marked point process technique. During the filament detection, we use the centre point of each pair instead of the positions of galaxies to avoid a built-in influence of pair orientation on the filament construction. For pairs lying within filaments (3012 cases), we calculate the angle between the line connecting galaxies of each pair and their host filament. To avoid redshift-space distortions, the angle is measured in the plain of the sky. }
   {The alignment analysis shows that the orientation of galaxy pairs correlates strongly with their host filaments. The alignment signal is stronger for loose pairs, with at least 25\% excess of aligned pairs compared to a random distribution. 
The alignment of galaxy pairs and filaments measured from the observational data is in good concordance with the alignment in the Millennium simulation and thus provides support to the $\Lambda$CDM formalism.}
   {}


   \maketitle
%
%________________________________________________________________

\section{Introduction}

The main driving force of structure formation in the expanding Universe is the gravitational collapse. The collapse amplifies any initial anisotropy of matter distribution. As first pointed out by \citet{Zeldovich:70}, the contraction is locally dominated by one dimension at a time. The speed of the collapse depends on the local density and the initial amplitude of perturbations, thus in numerical simulations, a complex pattern of nearly 2-dimensional sheets, nearly 1-dimensional filaments and roughly spherical small density peaks (knots) develops, framing large under-dense void regions \citep[][etc]{Doroshkevich:80, Klypin:83, Davis:85, Gramann:93, Sheth:04}. Within this network, matter flows along well defined paths, with the voids feeding the sheets, the sheets feeding the filaments and the filaments finally channeling everything towards the knots \citep[][and references therein]{Cautun:14}.

Observational confirmation for the existence of a characteristic pattern of large-scale inhomogeneities came already decades ago \citep[][etc]{Joeveer:78a, Joeveer:78b, Peebles:80, deLapparent:86, Bond:96}. However, the low number density of sufficiently luminous galaxies sets severe limits to the possibilities of comparing the actual structures to the simulated ones. As a result, the fine structure and the dynamics of the actual cosmic web can be directly probed only within our local volume, which may not be representative of the Universe at larger scales. Nevertheless, the existing peaces of evidence about the dynamics within our super-galactic neighbourhood support the void-sheet-filament-cluster sequence of matter flows \citep{Tully:14,Tempel:14d,Karachentsev:14}.

As an alternative to the simple mapping of structures, a method for testing and constraining structure formation scenarios is provided by the alignment of structures. Anisotropic collapse and tidal torques should induce spinning up and stretching out of dark matter haloes along the general velocity field within a large-scale structure element \citep{Codis:12,Aragon-Calvo:14,Libeskind:14,Welker:14}. Observationally, corresponding hints for some alignment between various structures have been found: elliptical galaxies and satellite systems around massive galaxies tend to be elongated in the direction of galactic filaments \citep{Tempel:13a,Lee:15,Zhang:15}, while galaxy discs tend to be oriented perpendicular to filaments \citep{Tempel:13a,Tempel:13}. The former findings can be explained with galaxy mergers and satellite accretion occurring along the filaments, while the latter effect can be reconciled with matter infall from voids and sheets around the filaments \citep{Dubois:14}.

In this paper we take a new step in the alignment probing by measuring the angle between the orientation of galaxy pairs and the orientation of their host filaments. The pairs and filaments have been extracted from the Sloan Digital Sky Survey (SDSS). For filament detection we use the Bisous model, which is based on a marked point process \citep{Tempel:14a}. The Bisous filaments have already proven to be handy for probing the galaxy content and other properties of the cosmic web \citep[e.g.][]{Tempel:14b, Tempel:14c, Guo:15}.

Throughout the paper we assume the Wilkinson Microwave Anisotropy Probe (WMAP) cosmology: the Hubble constant $H_0 = 100\,h\ \mathrm{km\,s^{-1}Mpc^{-1}}$, with $h=0.697$, the matter density $\Omega_\mathrm{m}=0.27$ and the dark energy density $\Omega_\Lambda=0.73$ \citep{Komatsu:11}.

%__________________________________________________________________

\section{Data and methods}

\subsection{SDSS galaxy and group samples}

This work is based on the SDSS data release~10 \citep{York:00, Ahn:13}. We use the galaxy and group samples as compiled in \citet{Tempel:14}, which cover the main contiguous area of the survey (the Legacy Survey). Galaxy groups are extracted using the conventional friend-of-friend (FoF) grouping algorithm with a varying linking length \citep[see][for more details]{Tempel:14}. The flux-limited catalogue (apparent $r$-band magnitude limit is 17.77) extends to the redshift 0.2 (820~Mpc) and includes 588193 galaxies and 82458 groups with two or more members.

In \citet{Tempel:14} the redshift-space distortions of galaxy groups are suppressed using the rms sizes of groups in the plane of the sky and their rms radial velocities as described in \citet{Liivamagi:12}. Galaxies in groups are ascribed new radial distances in order to make the spatial distribution of galaxy groups approximately spherical. Such a compression removes the artificial line-of-sight filament-like structures and was necessary to retrieve a realistic filamentary pattern in the cosmic web.

\begin{figure}
   \centering
   \includegraphics[width=80mm]{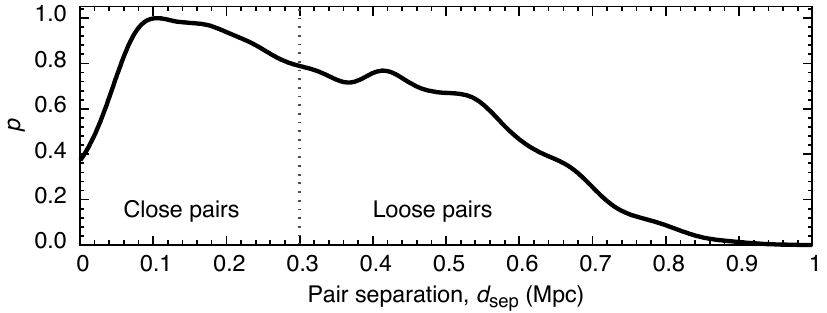}
   \caption{The distribution of pair separations in the plane of the sky. The vertical dashed line splits the sample into two equal subsamples.}
   \label{fig:pdf_sep}
\end{figure}

For our analysis we use only galaxy pairs (galaxy groups with two members in the \citet{Tempel:14} catalogue) that are located within filaments (see Sect.~\ref{sect:filsample} for the filament catalogue description). We consider a galaxy pair to be located in a filament if the geometrical centre point of the pair is closer than 0.5~Mpc to the filament axis. In total, this limits our sample to 3012 pairs. To avoid redshift space effects, the separation between the galaxies in a galaxy pair is measured in the plane of the sky. The distribution of pair separations for pairs located in filaments is shown in Fig.~\ref{fig:pdf_sep}.

Due to fibre collision in SDSS, some of the pairs in our sample are actually groups with three or more members. According to \citet{Tempel:12} the fraction of pairs that are affected by fibre collision is about 5\%, hence, we can ignore this effect in the current analysis.

\subsection{Semi-analytic galaxies in simulation}

For a quick interpretation of the results, we use the distribution of galaxies in the Millennium simulation \citep{Springel:05}, in which dark matter haloes are populated with galaxies according to a semi-analytical galaxy formation model \textsc{galform} \citep{Bower:06}. The simulated galaxy data and the galactic filaments are the same as used in \citet{Guo:15} and \citet{Tempel:15} -- we refer to these papers for a more detailed description. For the analysis, we only have considered galaxies brighter than $M_r=-19$. The galaxy pairs (groups with two members) were extracted using the FoF algorithm with a linking length 0.5~$h^{-1}\mathrm{Mpc}$. The pairs are separated into equal-size close and loose pair samples according to the real separation between galaxies. The limiting distance is 0.2~$h^{-1}\mathrm{Mpc}$ (i.e. roughly the same as for SDSS).

\subsection{Galaxy filament sample for the SDSS data}
\label{sect:filsample}

The detection of filaments is performed by applying an object/marked point process with interactions \citep[the Bisous process;][]{Stoica:05} to the spatial distribution of galaxies. This algorithm provides a quantitative classification that complies with the visual impression of the cosmic web and is based on a robust and well-defined mathematical scheme. More detailed descriptions of the Bisous model can be found in \citet{Stoica:07,Stoica:10} and \citet{Tempel:14a}. For reader convenience, a brief and intuitive summary is provided below.

Our model approximates the filamentary web by a random configuration of small segments. We assume that locally, galaxy conglomerations can be probed with relatively small cylinders, which can be combined to trace a filament if the neighbouring cylinders are oriented similarly. An advantage of the approach is that  it relies directly on the positions of galaxies and does not require any additional smoothing for creating a continuous density field.

The solution provided by our model is stochastic. Therefore, we find some variation in the detected patterns for different Markov chain Monte Carlo (MCMC) runs of the model. On the other hand, thanks to the stochastic nature of the method we gain a morphological and a statistical characterisation of the filamentary pattern simultaneously.

In practice, after fixing an approximate scale of the filaments, the algorithm returns the filament detection probability field together with the filament orientation field. Based on these data, filament spines are extracted and a filament catalogue is built in which every filament is represented by its spine: a set of points that define the axis of the filament.

Our spine detection follows two ideas. First, filament spines are located at the highest density regions outlined by the filament probability maps. Second, in these regions of high probability for the filamentary network, the spines are oriented along the orientation field of the filamentary network. See \citet{Tempel:14a} for more details of the procedure.

For the current study, we have rebuilt the filament catalogue in order to eliminate the potential influence of galaxy pairs to the filament orientation. For filament extraction, each galaxy pair is replaced with a single point (the centre point of the two galaxies in the 3D space). Doing so, the orientation of a galaxy pair cannot affect the filament detection and the orientations of filaments and galaxy pairs are measured independently.

The method and parameters for filament extraction are exactly the same as in \citet{Tempel:14a}. The assumed scale (radius) of the filaments is roughly 0.7~Mpc\footnote{Most of the galaxies are close to the filament axis and 90\% of galaxies inside 0.7~Mpc are closer than 0.5~Mpc to the filament axis.}. Because the survey is flux-limited, the sample is very diluted farther away, hence we are able to detect filaments of this scale only up to the redshift 0.15 ($\approx 640$~Mpc). To reduce the survey edge effects, we use only filaments that are farther than 10~Mpc from the survey boundary. In addition, to restrict ourselves with reliable filaments we dismiss all  filaments shorter than 10~Mpc.

\subsection{Measuring the alignment signal}

As explained above, the determination of the orientation of filaments and the orientation of galaxy pairs are independent by construction and we do not have to worry about an intrinsic alignment caused by the method. However, the orientation of galaxy pairs and filaments are affected by the redshift-space distortions. To avoid any systematic effects introduced by the redshift-based distances, we measure the alignment in the plane of the sky. For that we project galaxy pairs and filaments to the plane of the sky and measure the angle between filament orientation and pair orientation. The measured angle is restricted to be in the range of 0 to 90 degrees. We calculate the alignment signal using a kernel density estimation.\footnote{In Appendix~A of \citet{Tempel:14b} we show that the kernel density estimation is better than a simple histogram and is more representative of the underlying probability distribution.}

To estimate confidence regions of the measured alignment signal, we generate 10000 Monte Carlo samples with the same number of pairs as in the original sample and conduct the same alignment analysis. In the figures below, we show the 95\% confidence regions for this randomised signal. The Kolmogorov-Smirnov test (KS-test) probabilities that the angle distributions are drawn from a uniform distribution are also shown. To compare the two measured alignments, we use the KS-test to calculate the probability that the distributions are drawn from the same distribution. The KS-test values are indicated in the figures.

\section{Results}

\begin{figure}
   \centering
   \includegraphics[width=76mm]{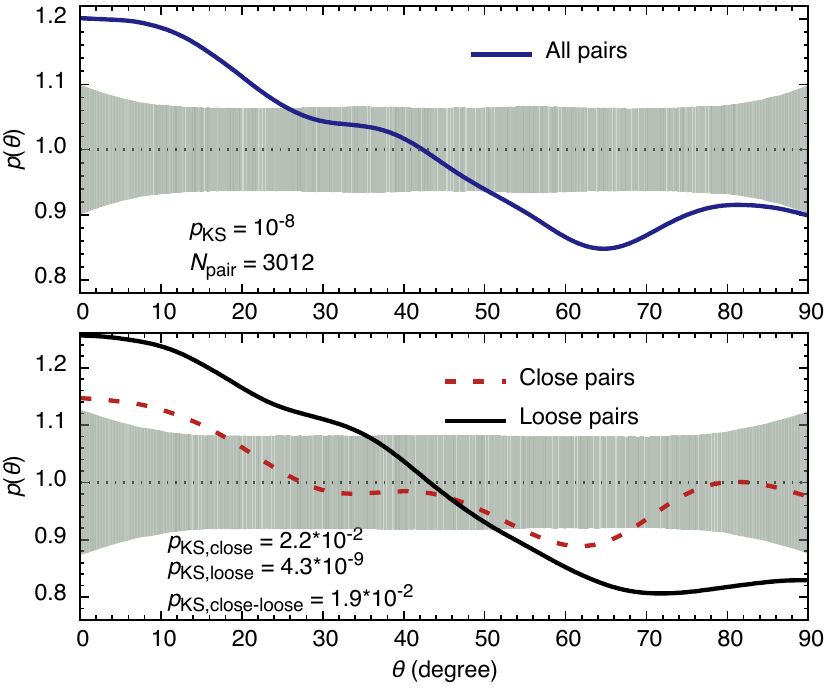}
   \caption{\emph{Upper panel}: probability distribution function (blue solid line) of the projected (in the plane of the sky) angles between galaxy pairs and their host filaments. The KS-test value that the sample is drawn from a uniform distribution is $10^{-8}$. The filled area shows the 95\% confidence region for a randomised distribution of 3012 pairs. \emph{Lower panel:} the same as in the upper panel for two equal-size subsamples: close pairs ($d_\mathrm{sep}<0.3$~Mpc; red dashed line) and loose pairs ($d_\mathrm{sep}>0.3$~Mpc; black solid line).}
   \label{fig:full}
\end{figure}

Figure~\ref{fig:full} (upper panel) shows the probability distribution of the angles between galaxy pair orientations and their host filament orientations. As explained above, the angles are measured in the plane of the sky in order to avoid possible effects of redshift-space distortions. It is seen that galaxy pairs are strongly aligned with galactic filaments: the number of galaxy pairs aligned with the filament spine within $\pm$ 10 degrees is about 20\% higher than would be in the case of a random distribution, and the pair number decreases nearly monotonically when moving towards perpendicularity with the filament spine. The KS-test probability that the alignment angles are drawn from a uniform distribution is $p_\mathrm{KS}=10^{-8}$ ($\sim 6.5\sigma$ signal).

In the lower panel of Fig.~\ref{fig:full} we have split the pair sample into two equal-size subsamples according to the projected distance between the galaxies of each pair. The demarcation between close and loose pairs (0.3~Mpc) is indicated in Fig.~\ref{fig:pdf_sep}. Figure~\ref{fig:full} shows that galaxy pair alignment with galactic filaments is much stronger for loose pairs than for close pairs, while the alignment significance remains above 95\% ($2\sigma$) in both cases. The KS-test probability that the close and loose pair samples are drawn from the same underlying distribution is 0.019, hence, the close and loose samples should be intrinsically different. However, the close-pair sample is contaminated by loose pairs oriented close to the line of sight, raising the measured amount of non-aligned pairs.

\begin{figure}
   \centering
   \includegraphics[width=76mm]{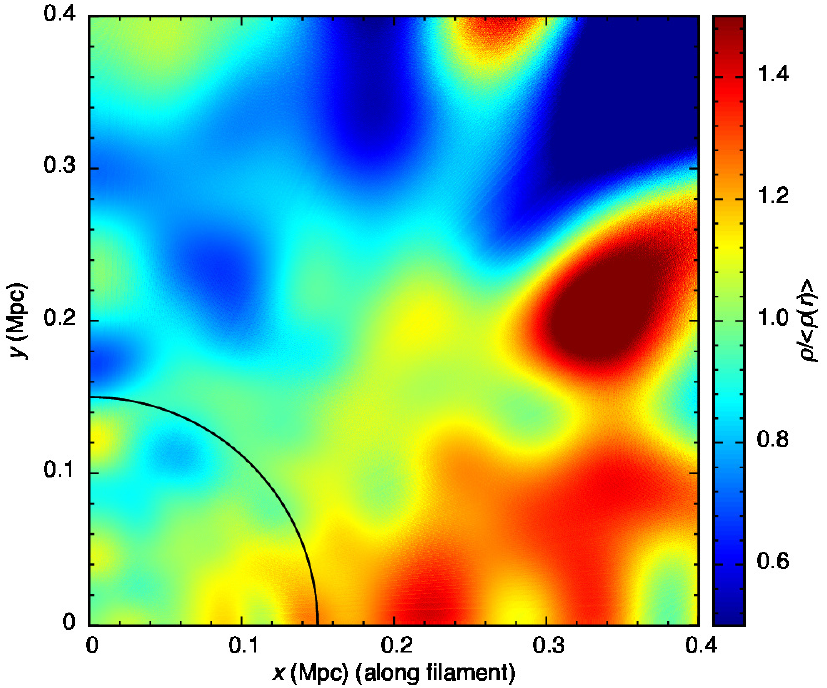}
   \caption{The distribution of stacked galaxy pairs in the plane of the sky. Galaxy pairs are stacked based on the orientation of the host filaments (represented by the $x$-axis); the centre point of each pair matches with the origin of the coordinate axes. The densities are shown in units of spherically averaged mean number densities. The black solid line shows the demarcation of close and loose pairs.}
   \label{fig:map}
\end{figure}

We have checked also for the effect of sample splitting on the basis of various other measurable quantities: galaxy luminosity, colour, magnitude difference, distance, morphology, distance from filament axis, etc. In some cases, a small impact on the alignment signal was noticeable, but statistically insignificant. In principle, the statistical significance for the alignment signal can be increased either by increasing the sample size or by measuring the alignment in the real space rather than in projection. The latter requires a proper treatment of redshift-space distortions, which is not a straightforward procedure and is beyond the scope of the present analysis.

To visualise the galaxy pair alignment in filaments, we show a stacked distribution of the galaxy pairs in the plane of the sky in Fig.~\ref{fig:map}. For the stacking, the centre point of pairs was fixed at the origin of the coordinate axes and all filaments were oriented along the $x$-axis. The figure shows that for the loose pairs, the alignment occurs uniformly across the whole range of separations. Note that the apparent density maximum in this figure around $[x=0.35, y=0.2]$ is actually statistically insignificant -- it corresponds to an enhancement of the random distribution by just a few galaxies at the given projected distance. 

\begin{figure}
   \centering
   \includegraphics[width=76mm]{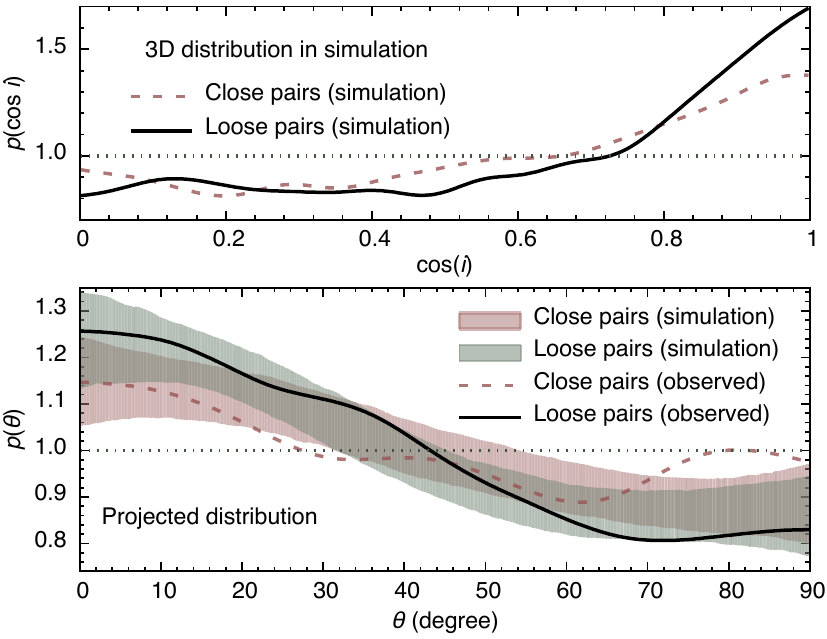}
   \caption{\emph{Upper panel:} the alignment of close- and loose galaxy pairs with the host filaments in the Millennium simulation; $i$ denotes the real-space angle between a filament and a galaxy pair. \emph{Lower panel:} the angle distribution for close and loose pairs in the SDSS data (solid and dashed lines; same as in Fig.~\ref{fig:full}) compared to the sky projection of a model distribution, in which the real-space alignment between pairs and filaments of the Millennium simulation (shown in the \emph{upper panel}) is ascribed to the actual SDSS filaments. The modelled distributions are shown as filled regions corresponding to the 95\% confidence level.}
   \label{fig:model}
\end{figure}

While the detected 2D alignment indicates that also in the 3D space, a significant fraction of galaxy pairs has to be aligned with filaments, a quantification and a more detailed interpretation of the measured signal is not straightforward due to projection effects and redshift distortions. Instead, we can check whether the measured signal corresponds to the predictions of the $\Lambda$CDM formalism. To this end, using data from the Millennium simulation (as described above), we have studied the pair alignment with host filaments in the real space, splitting the pair sample into close and loose pairs according to the real-space distance\footnote{The use of real-space distances guarantees that the dependence on pair separation is intrinsic and not affected by projection effects. However, when splitting the sample based on the projected distance, the measured alignment signals remain roughly the same.}. The upper panel of Fig.~\ref{fig:model} shows the resultant real-space alignment with the filaments in the simulation data. As in the case of the SDSS data, both subsamples show significant alignment, with close pairs being somewhat less aligned than the loose ones. For a more direct comparison with the SDSS pairs, we have applied the alignment of the Millennium pairs to the SDSS filaments. In 2D projection, the resultant angle distribution is shown in the lower panel of Fig.~\ref{fig:model} as filled regions. For loose pairs, the pair alignment of the simulation matches well with the observed one. For close pairs, the distributions nearly match within the 95\% confidence level, except for angles beyond about 70 degrees, where the observed pairs seem to exhibit a population perpendicular to the host filaments. However, the mismatch can be a result of the sample variance and does not provide grounds for invalidating the simulation.

\section{Conclusions and implications}

We used the SDSS Legacy survey to analyse the alignment between galactic filaments and galaxy pairs. The filamentary pattern in galaxy distribution was detected using the Bisous model and a catalogue of galactic filaments was constructed. To overcome possible contamination by galaxy pair orientations, the galaxy pairs were treated as single points during the filament extraction. The alignment is measured in the plane of the sky to avoid redshift-space distortions.

The data show a clear tendency of galaxy pairs to align with their host filaments. The alignment is stronger for loose pairs and weaker for close pairs. The detected alignment is statistically highly significant, reaching up to $6.5\sigma$. Quantitatively, the alignment is similar to the alignment measured for galaxy pairs and filaments extracted from the Millennium simulation, thus being in concordance with the $\Lambda$CDM paradigm.

Recently, the alignment of satellite galaxies and the cosmic web structures have been analysed by \citet{Lee:15} and \citet{Tempel:15}. Both studies conclude that the distribution of satellites around primary galaxies is elongated in the direction of galactic filaments.

A simulation-based test carried out by \citet{Tempel:14c} showed that the given filament detection method also recovers the underlying velocity field. We can thus deduce that the alignment of galaxy pairs and satellite systems with their host filaments is a direct result of the anisotropic velocity field, leading towards a preferred direction of satellite accretion and gravitational collapse.

\begin{acknowledgements}
     We thank referee for useful suggestions and comments. We thank our colleague Enn Saar for stimulating discussions. This work was supported by institutional research funding IUT26-2, IUT40-2 and TK120 of the Estonian Ministry of Education and Research. Funding for SDSS-III has been provided by the Alfred P. Sloan Foundation, the Participating Institutions, the National Science Foundation, and the U.S. Department of Energy Office of Science. The SDSS-III web site is http://www.sdss3.org/.
\end{acknowledgements}

%\bibliographystyle{aa}
%\bibliography{mybib}{}

\end{document}